\documentclass[aps,prd,twocolumn,groupedaddress,showpacs]{revtex4}

\usepackage{graphicx}
\usepackage{dcolumn}
\usepackage{bm}
\usepackage{amsmath}
\usepackage{epstopdf}
\usepackage{amsfonts}
\usepackage{amssymb}
\usepackage{xcolor}
\usepackage[utf8]{inputenc}
\usepackage{hyperref}
\providecommand{\U}[1]{\protect\rule{.1in}{.1in}}

\newcommand{\baa}{\begin{align}}
\newcommand{\eaa}{\end{align}}

\newcommand{\be}{\begin{equation}}
\newcommand{\ee}{\end{equation}}
\newcommand{\bea}{\begin{eqnarray}}
\newcommand{\eea}{\end{eqnarray}}

\begin{document}



\title{Quasinormal modes of an improved Schwarzschild black hole}


\author{\'Angel Rinc\'on}
\affiliation{
Instituto de F\'isica, Pontificia Universidad Cat\'olica de Valpara\'iso,
\mbox{Avenida Brasil 2950, Casilla 4059, Valpara\'iso, Chile.}}
\email{angel.rincon@pucv.cl}

\author{Grigoris Panotopoulos}
\affiliation{Centro de Astrof\'{\i}sica e Gravita{\c c}{\~a}o, Departamento de F{\'i}sica, Instituto Superior T\'ecnico-IST, Universidade de Lisboa-UL, 
Av. Rovisco Pais, 1049-001 Lisboa, Portugal.}
\email{grigorios.panotopoulos@tecnico.ulisboa.pt}

%

\date{\today}

\begin{abstract}
We compute the quasinormal frequencies for scalar and electromagnetic perturbations of an improved Schwarzschild geometry in the framework of asymptotically safe gravity, which is one of the approaches to quantum gravity. Adopting the widely used WKB semi-classical approximation, we investigate the impact on the spectrum of the angular degree, the overtone number as well as the black hole mass. We summarize our numerical results in tables, and for better visualization, we show them graphically as well. All modes are found to be stable. Finally, we compare our numerical results with those corresponding to the classical Schwarzschild solution as well as to the results obtained using a different approach. Our findings show that i) a different cut-off identification does not affect the spectra significantly, and ii) for hypothetical objects with masses comparable to the Planck mass, the difference in the numerical values between the modes of the classical solution and the modes of the improved solution studied here is of the order of a few per cent. On the contrary, for realistic, astrophysical BHs no difference in the frequencies is observed.
\end{abstract}


\maketitle


\section{Introduction}\label{Intro}

Black holes (BHs), a generic prediction of Einstein's General Relativity (GR) and of other metric theories of gravity, and at the same time the simplest objects in the Universe, are of paramount importance both for classical and quantum gravity. Hawking's radiation \cite{hawking1,hawking2}, although it has not been detected yet, has always been attracted a lot of attention in the community. What is more, the by now numerous direct detections of gravitational waves from black hole mergers \cite{ligo1,ligo2,ligo3,ligo4,ligo5}, combined with the first image of the supermassive black hole \cite{L1,L2,L3,L4,L5,L6} located at the centre of the giant elliptical galaxy Messier 87 by the Event Horizon Telescope project \cite{project} last year, have boosted even more the interest in studying several different aspects of black hole physics.

\smallskip

Isolated black holes are ideal objects. In practice, realistic black holes in Nature are never isolated. Instead, they are in a constant interaction with their environment. When a black hole is perturbed, it responds by emitting gravitational waves. Quasinormal modes (QNMs) are characteristic frequencies with a non-vanishing imaginary part, which encode the information on how black holes relax after the perturbation has been applied. The QN frequencies depend on the details of the geometry and on the type of the perturbation (scalar, vector, tensor or fermionic), but not on the initial conditions. Black hole perturbation theory \cite{wheeler,zerilli1,zerilli2,zerilli3,moncrief,teukolsky} and QNMs become relevant during the ring down phase of a black hole merger, the stage where a single distorted object is formed, and where the geometry of space time undergoes dumped oscillations due to the emission of gravitational waves. Thanks to gravitational wave Astronomy we have now a powerful tool at our disposal to test gravitation under extreme conditions. For excellent reviews on the topic see \cite{review1,review2,review3}, and also the Chandrasekhar's monograph \cite{chandra}, which is the standard textbook on the mathematical aspects of black holes. 

\smallskip

It is well-known that the formulation of a consistent quantum theory of gravity is still an open task in modern theoretical physics. Although as of today several approaches to the problem do exist in the literature (for a partial list see e.g. \cite{QG1,QG2,QG3,QG4,QG5,QG6,QG7,QG8,QG9} and references therein), there is one property in particular that all of those have in common. Namely, the basic parameters that enter into the action defining the model at hand, such as Newton's constant, electromagnetic coupling, the cosmological constant etc, become scale dependent quantities. This does not come as a surprise of course, since scale dependence at the level of the effective action is a generic feature of ordinary quantum field theory. In theories of gravity, the scale dependence is expected to modify the horizon, the thermodynamics as well as the QN spectra of classical black hole backgrounds \cite{Koch:2016uso,Rincon:2017ypd,Rincon:2017goj,Rincon:2017ayr,
Contreras:2017eza,Rincon:2018sgd,Contreras:2018dhs,Rincon:2018lyd,
Rincon:2018dsq,Contreras:2018gct,Rincon:2019cix,Contreras:2019fwu,
Contreras:2019cmf,Contreras:2018gpl}. Also, the Sagnac effect \cite{Rincon:2019zxk}, the evolution of trajectories of photons \cite{Fathi:2019jid}, some cosmological solutions \cite{Hernandez-Arboleda:2018qdo,Canales:2018tbn}, and transverse wormhole solutions \cite{Contreras:2018swc} have also been investigated.

\smallskip

In the context of theories beyond classical GR, apart from the aforementioned approach based on scale-dependent gravity, there is yet another method, which is usually called improvement asymptotically safe (AS) gravity \cite{Bonanno:2000ep,Bonanno:2001xi,Reuter:2003ca}. Also, for RG-improved cosmologies and inflationary models from asymptotic safety see e.g. \cite{Platania:2020lqb,Rubano:2004jq,Bonanno:2002zb,Bonanno:2001hi,Liu:2018hno,Hindmarsh:2012rc}, and for recent progress \cite{Platania:2019kyx,Moti:2018uvl,Koch:2013owa,Bonanno:2016dyv,
Pawlowski:2018swz,Gonzalez:2015upa}.

\smallskip

In that scenario, the main idea is to integrate the beta function for the gravitational coupling to compute Newton's constant as a function of some energy scale $k$, $G(k)$.  After that, Newton's constant is inserted into the classical black hole solution and the improved lapse function is obtained. It is essential to notice that the gravitational coupling depends on some arbitrary renormalization scale $k$ (see next section). Finally, a link between the energy scale $k$ and the radial coordinate $r$ must be established. It is only after that final step that the complete solution for an improved black hole is obtained. Those extended solutions, inspired by the asymptotic safety program, are expected to modify the classical black hole solutions incorporating quantum features.

\smallskip

Given the importance and relevance of the QNMs of black holes in gravitational wave Astronomy, it would be interesting to see what kind of spectra are expected for improved black holes. In the present work we propose to compute the QN frequencies for scalar as well as for electromagnetic perturbations of an improved Schwarzschild black hole in the framework of asymptotically safe gravity. It should be mentioned here that in a couple of previous works the QN spectra of improved BHs have been considered \cite{chinosII,Li:2013kkb,Konoplya:2019xmn}. Our work differs from those in the following: First, in \cite{chinosII,Li:2013kkb} the lapse function, which is given by an approximate expression, is different than ours. Moreover, in \cite{Konoplya:2019xmn} the author computed the QN spectra for scalar, Dirac and electromagnetic perturbations of an improved black hole in the approach suggested by Kozakov and Solodukhin \cite{Kazakov:1993ha}. In this approach 
renormalizability was provided neglecting non-spherical deformations, while at the same time it seems  to be the only approach for which the problem was solved exactly and non-perturbatively \cite{Konoplya:2019xmn}. 

\smallskip

The plan of our work is the following: In the next section we review an improved Schwarzschild black hole, while in the third section we discuss the wave equation with the corresponding potential barrier for the scalar as well as electromagnetic perturbations. In the fourth section we compute the quasinormal frequencies adopting the WKB approximation of 6th order, and we discuss our results. Finally, we summarize our work with some concluding remarks in section five. We adopt the mostly positive metric signature $(-,+,+,+)$, and we work in geometrical units where the universal constants are set to unity, $c=1=G_0$.

\section{Renormalization group-improvement black hole solutions} \label{Classical}
\noindent

In this section we shall review the improved Schwarzschild solution obtained for the first time by Bonanno and Reuter \cite{Bonanno:2000ep}, following closely their notation.

\smallskip

\begin{figure*}[ht]
\centering
\includegraphics[width=0.32\textwidth]{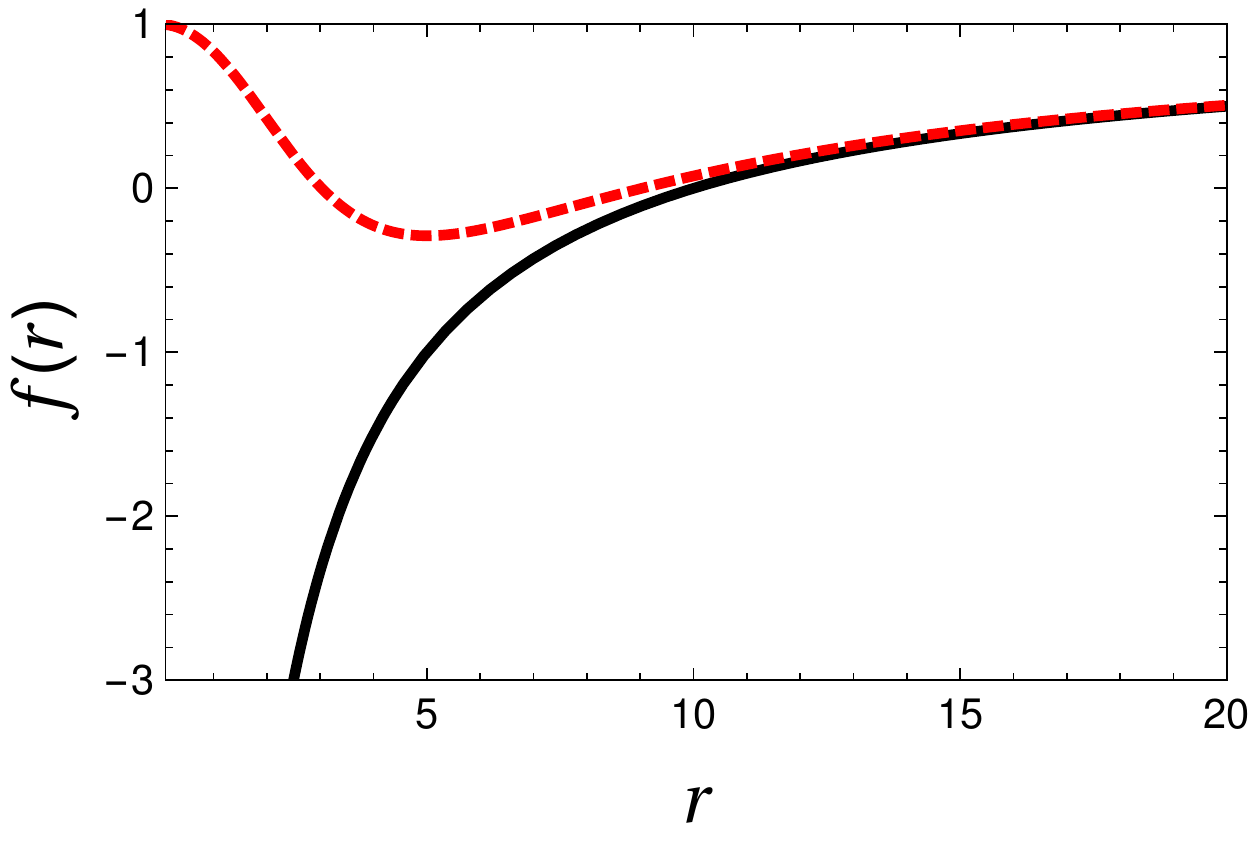}   \
\includegraphics[width=0.32\textwidth]{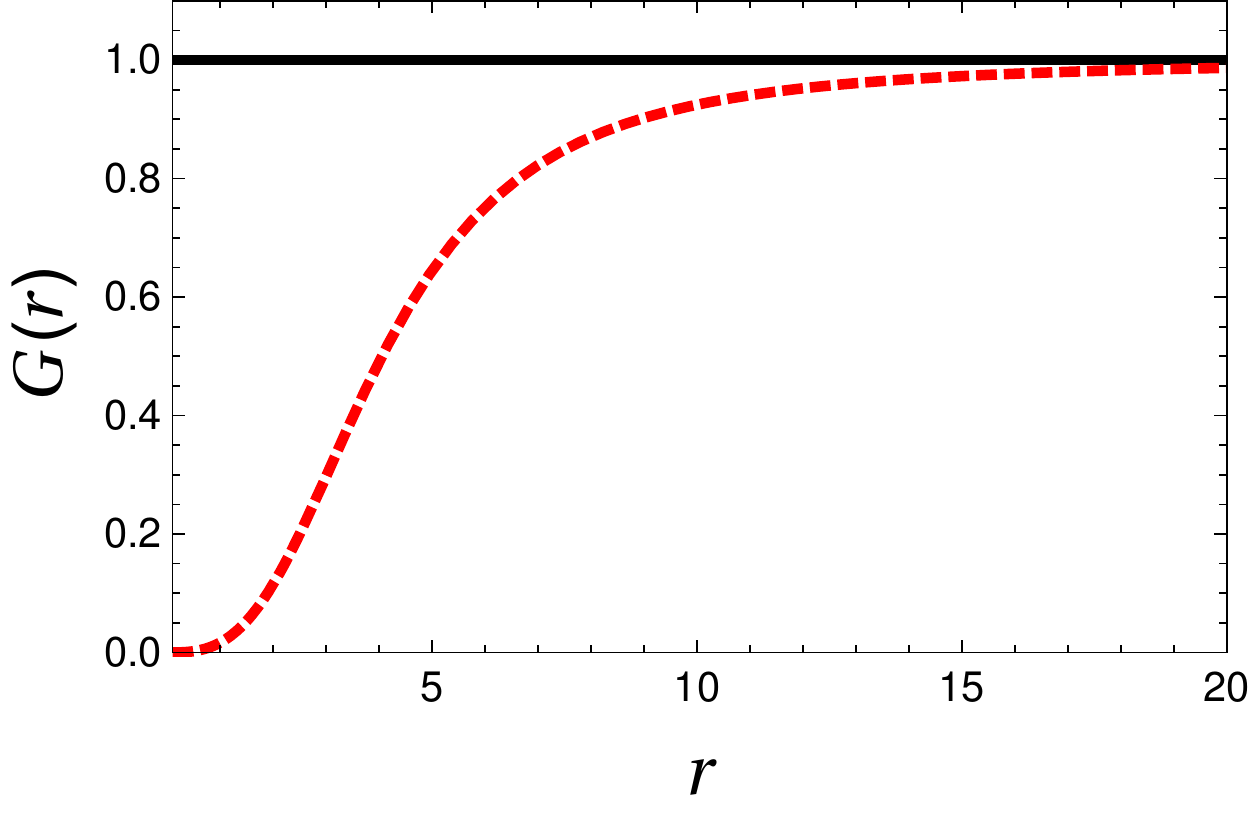}   \
\includegraphics[width=0.33\textwidth]{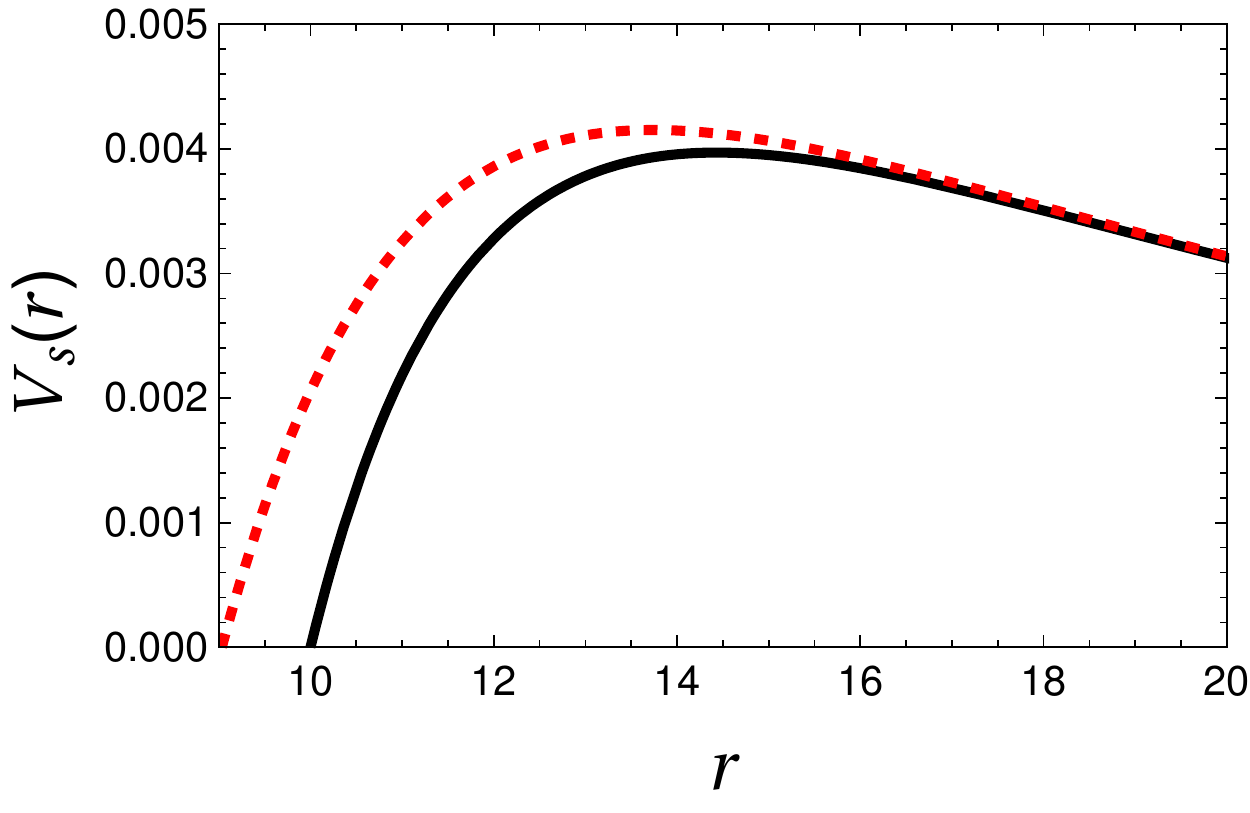}   \
\caption{
Left, middle and right panels show the lapse function $f(r)$, the gravitational coupling $G(r)$ and the effective potential barrier for scalar perturbations $V_s(r)$, respectively, for $M=5$ and $l=1$. Shown are: i) the classical case ($\tilde{\omega} = 0$) in solid black line, and ii) the improved BH solution considered in this work ($\tilde{\omega}=118/15 \pi$ and $\gamma=9/2$) in dashed red line.
}
\label{fig:report}
\end{figure*}

The starting point is the average Einstein-Hilbert action, avoiding ghosts, in four dimensional spacetime :
\begin{align}
\Gamma_k[\mathfrak{g}] &= \frac{1}{16 \pi G(k)} \int \mathrm{d}^4 x \sqrt{\mathfrak{g}} \Bigl( - R(\mathfrak{g}) + 2 \bar{\lambda}(k)\Bigl),
\end{align}
where $G(k)$ and $\bar{\lambda}(k)$ are the running Newton's constant and the cosmological constant, respectively, and $\mathfrak{g}$ is the corresponding metric.
Moreover, for the evolution of the scale-dependent couplings we use the well-known
Wetterich equation given by
\begin{align}
\partial_{t}\Gamma_{k} &= \frac{1}{2} \text{Tr}\left(\frac{\partial_{t}\mathcal{R}_{k}}{\Gamma^{(2)}_{k}[\phi] + \mathcal{R}_{k}}\right),
\end{align}
where $t=\ln(k)$, $\Gamma_{k}^{(2)}$ stands for the Hessian of $\Gamma_k$ with respect to $g_{\mu \nu}$, and $\mathcal{R}_k$ is a filtering function, as was pointed out in Ref.~\cite{Bonanno:2000ep}. To be more precise, $\mathcal{R}_k(p^2) \propto k^2 R^{(0)}(z)$, being $z \equiv p^2/k^2$. To carry out the calculation, it is required to choose a certain arbitrary smooth function $\mathcal{R}_{k}$. 
For practical reasons, usually the form of $R^{(0)}$ is fit as a exponential function, namely
\begin{align}
R^{(0)}(z) \equiv \frac{z}{e^z - 1}.
\end{align}
Inserting $\Gamma_k$ into the Wetterich equation and projecting the flow onto the subspace spanned by the Einstein-Hilbert truncation, one obtains a system of coupled differential equations for the dimensionless Newton's constant $g(k)$, which satisfies:
\begin{align}\label{gkk}
g(k) &\equiv k^{2}G(k). 
\end{align}
%
%
The evolution of the corresponding dimensionless gravitational coupling is given by
\begin{align} \label{betag}
\frac{\mathrm{d}g(t)}{\mathrm{d}t} \equiv \beta(g(t)) = \left[ 2 + \frac{B_1 g(t)}{1- B_2 g(t)} \right] \: g(t),
\end{align} 
where $B_1$ and $B_2$ are defined in \cite{Bonanno:2000ep}. 
In particular, they are
\begin{align}
B_1 \equiv B_1(0) &= -\frac{1}{3\pi} \bigg( 24 \Phi^2_2(0) - \Phi_{1}^{1} (0) \bigg),
\\
B_2 \equiv B_2(0) &= \frac{1}{6\pi} \bigg( 18 \tilde{\Phi}^2_2(0) -5 \tilde{\Phi}_{1}^{1} (0) \bigg),
\end{align}
where the two auxiliary functions are defined by
\begin{align}
\Phi^{p}_{n} (w) &\equiv \frac{1}{\Gamma(n)} 
\int_{0}^{\infty} \mathrm{d}z \ z^{n-1}
\frac{ R^{(0)}(z)  - z {R^{(0)}}'(z) }{ (z + R^{(0)}(z) + w)^{p} },
\\
\tilde{\Phi}^{p}_{n} (w) &\equiv \frac{1}{\Gamma(n)} \int_{0}^{\infty} 
\mathrm{d}z \ z^{n-1}
\frac{ R^{(0)}(z) }{ (z + R^{(0)} (z) + w)^{p} }.
\end{align}
Notice that Eq.~\eqref{betag} can be integrated to obtain an explicit form for $g(k)$, and using Eq.~\eqref{gkk}, the dimensionfull Newton's coupling is found to be
\begin{align} \label{basic}
G(k) &= \frac{G_0}{1 + \tilde{\omega} G_0 k^2}.
\end{align}
Therefore, deviations from the classical solution are important at high energy scales, while in the opposite limit the classical space-time is recovered. This is graphically shown in the left and middle panels of Fig.~\ref{fig:report}. For the time being we keep $G_0$ in the expressions, and only in the end we shall set it to unity. 

Up to now, the procedure is restricted to taking a particular action and using the Wetterich equation to obtain the beta function for the gravitational coupling. But the crucial point in the ``improved" approach is the connection of $k$ with some physical parameter.

Recently, identifications involving the Kretschmann scalar have been used (see for instance \cite{Held:2019xde,Platania:2019kyx,Pawlowski:2018swz}). Also notice that a simple RG-improvement with identification of $k$ with the Kretschmann scalar produces an Hayward type of solution \cite{Held:2019xde}, improving for the second time the RG-improved solution leading to a singular solution \cite{Pawlowski:2018swz} and, therefore, performing a self-consistent RG-improvement procedure leads to a Dymnikova black hole \cite{Platania:2019kyx}.

\smallskip

In the following, however, we will use one possible identification, implemented some time ago in black hole physics. Thus we take
\begin{align} \label{distance_k}
k(P) = \frac{A}{d(P)},
\end{align}
where $A$ is a numerical factor to be fixed, and $d(P)$ is the distance scale which provides the relevant cut-off for Newton's constant when the test particle is located at the point $P$ of the black hole space-time \cite{Bonanno:2000ep}. Here we shall employ the following function:
\begin{align} \label{kpaper}
k(r) \equiv \left( \frac{r^3}{ r + \gamma G_0 M }  \right)^{1/2} .
\end{align}
It is essential to point out that this identification is not chosen at random. The renormalization scale $k(r)$ is a modified proper distance: since Eq.~\eqref{distance_k} is singular at the horizon, Bonanno and Reuter invented this function which interpolates smoothly between the behaviour of the proper distance close to $r=0$ and at infinity.
Inserting \eqref{kpaper} into eq. (\ref{basic}) we finally obtain the explicit form for the gravitation coupling, which is found to be
\begin{align}
G(r) &= \frac{G_0 r^3}{r^3 + \tilde{\omega} G_0 
\left (  r + \gamma G_0 M \right) } , 
\end{align}
where the parameters $\gamma, \tilde{\omega}$ are found to be \cite{Bonanno:2000ep}
\begin{align}
\tilde{\omega} = \frac{118}{15 \pi},
\hspace{1cm}
\text{and}
\hspace{1cm}
\gamma = \frac{9}{2}.
\end{align}
The interested reader may consult \cite{Bonanno:2000ep} for more details explanation regarding those precise numerical values.

\smallskip

The implementation of those ideas for the Schwarzschild geometry may be immediately seen, and it is the following: The line element for the metric tensor in the usual Schwarzschild form is still given by
\begin{align}
\mathrm{d}s^2 &= -f(r)\mathrm{d} t^2 + f(r)^{-1} \mathrm{d} r^2 + r^2 \: \mathrm{d}\Omega^2,
\end{align}
where $r$ is the radial coordinate, and $\mathrm{d} \Omega^2 = \mathrm{d \theta^2 + sin^2 \theta \: d \phi^2}$ is the line element of the unit two-dimensional sphere, while the lapse function $f(r)$ is given by the usual expression for a classical Schwarzschild solution
\begin{align}
f(r) & = 1 - \frac{2 M G(r)}{r}.
\end{align}
the only difference being the fact that now Newton's constant depends on $r$. The metric function is finally computed to be (where now we set $G_0=1$)
\begin{align}
f(r) = 1-\frac{2 M r^2}{r^3 +  \tilde{\omega} (r + \gamma M)}
\end{align}
with $M$ being the mass of the black hole. 
This is the main result regarding the lapse function of this improved Schwarzschild black hole. 

\smallskip

\begin{figure*}[ht]
\centering
\includegraphics[width=0.48\textwidth]{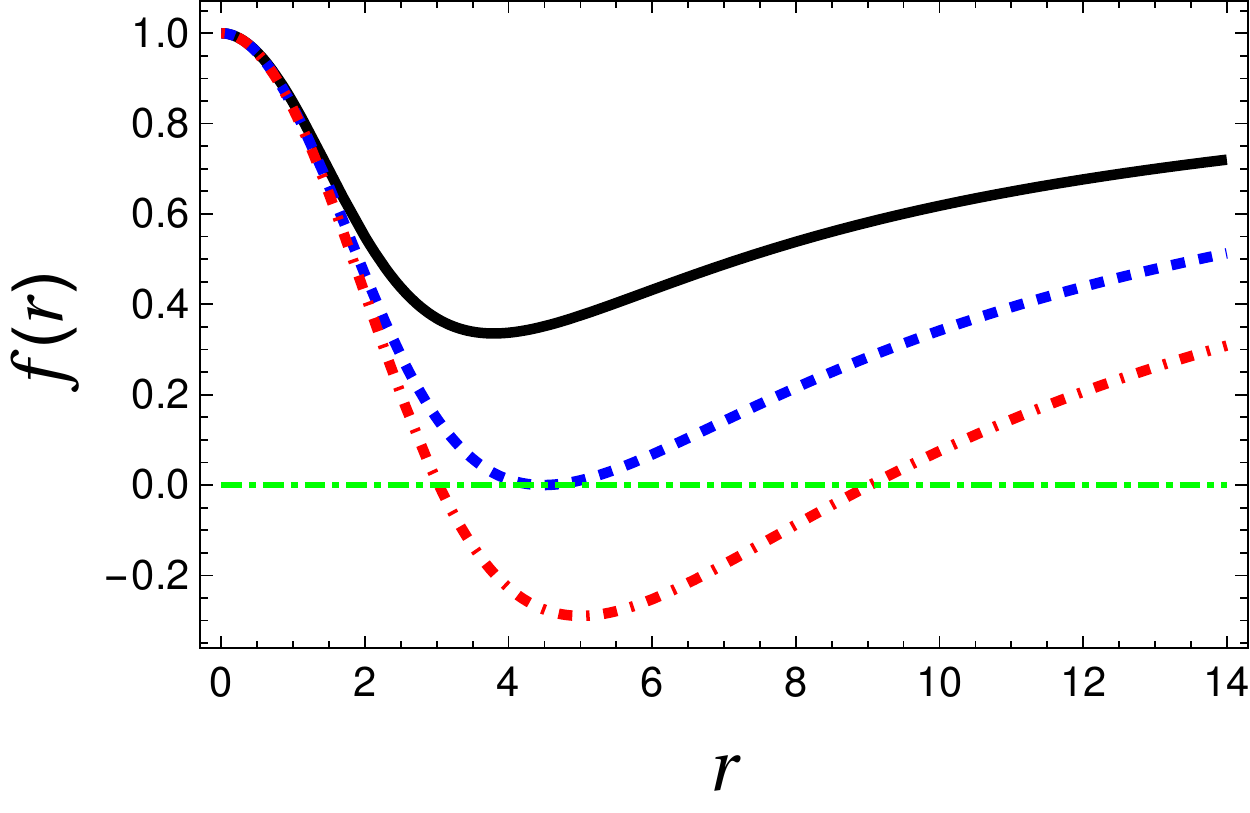}   \
\includegraphics[width=0.48\textwidth]{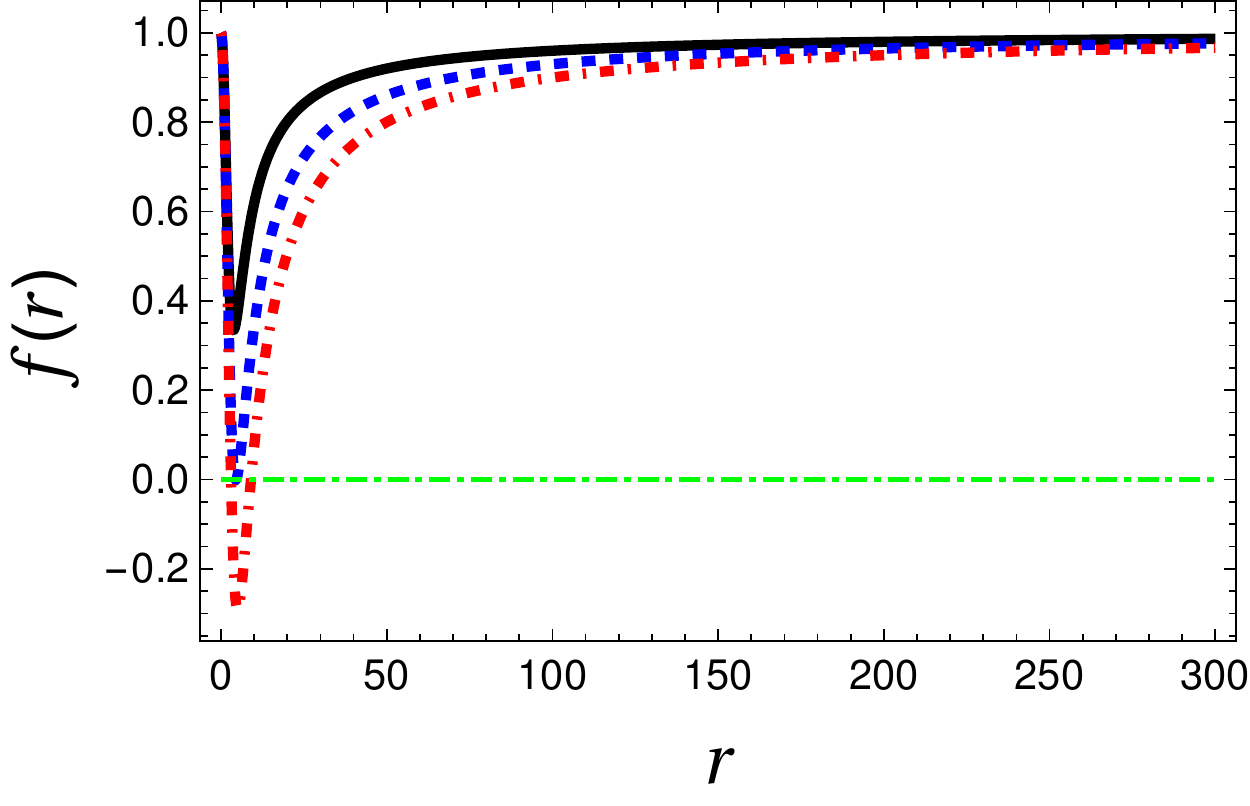}   \
\caption{
{\bf{LEFT:}} Improved lapse function $f(r)$ versus radial coordinate $r$ for a Schwarzschild black hole. We have assumed three different values of the classical black hole mass: i) $M=2$ (solid black line, no horizons), ii) $M \approx 3.503$ (dashed blue line, critical black hole), and iii) $M=5$ (dot-dashed red line) where an event horizon $r_H$ and a Cauchy horizon $r_-$ exist.
{\bf{RIGHT:}} Same as in the left panel, but for a much wider range for the radial coordinate $r$.
}
\label{fig:lapse}
\end{figure*}

It is instructive to consider the two limiting cases, namely a) at low energy scales ($r \rightarrow \infty$) 
\begin{align}
f(r \rightarrow \infty) \simeq 1 - \frac{2 M}{r}
\end{align}
and at high energy scales ($r \rightarrow 0$)
\begin{align}
f(r \rightarrow 0) \sim 1 - \frac{2 r^2}{\gamma \tilde{\omega}} + \mathcal{O}(r^3),
\end{align}
This is graphically shown in Fig.~\ref{fig:lapse}. Moreover, in the left panel of the same figure the impact of the parameter $M$ on the solution is shown, where we plot the corresponding improved lapse function versus the radial coordinate for three different values of the parameter $M$. According to the value of the mass of the black hole there are three well-defined cases: i) No horizons (when $M < M_{\text{crit}}$), ii) critical black hole (when $M = M_{\text{crit}} \simeq 3.503$) where a single horizon $r_H$ exists, and iii) usual black hole (when $M > M_{\text{crit}}$) where an event horizon $r_H$ and a Cauchy horizon $r_{-}$ exist. In the discussion to follow we will focus on the last case where $M > M_{\text{crit}}$. 

\smallskip

Since the choice for the cut-off is not unique, we mention here \cite{chinosII,Li:2013kkb} as an example of another work where a different improved Schwarzschild solution was studied. The lapse function used in \cite{chinosII,Li:2013kkb} is given by the approximate expression
\begin{equation}
f_{\text{appr}}(r) \simeq 1 - \frac{2Mr}{r^2 + \xi}
\end{equation}
and the classical solution is recovered for $\xi=0$.


\section{Schr{\"o}dinger-like equation for scalar and electromagnetic perturbations} \label{Wave}

In this section we introduce the necessary ingredients to deal with the computation of the quasinormal modes for scalar perturbations. To that end, let us consider the propagation of a probe massless minimally coupled scalar field $\Phi$ in a fixed gravitational background. Its wave equation is given by the usual Klein-Gordon equation, namely
\cite{coupling,p11,p12}
\begin{equation}
\frac{1}{\sqrt{-\mathfrak{g}}} \partial_\mu 
\Bigl(
\sqrt{-\mathfrak{g}} \mathfrak{g}^{\mu \nu} \partial_\nu
\Bigl) 
\Phi = 0
\end{equation}
In order to solve the previous equation, we apply the separation of variables making the usual ansatz:
\begin{equation}\label{separable}
\Phi(t,r,\theta, \phi) = e^{-i \omega t} \frac{\psi(r)}{r} Y_l^m (\theta, \phi)
\end{equation}
where $Y_l^m$ are the usual spherical harmonics, and $\omega$ is the frequency to be determined. 
Making the previous ansatz it is straightforward to obtain for the radial part a Schr{\"o}dinger-like equation
\begin{equation}
\frac{\mathrm{d}^2 \psi}{\mathrm{d}x^2} + [ \omega^2 - V(x) ] \psi = 0
\end{equation}
with $x$ being the tortoise coordinate, i.e.,
\begin{equation}
x  \equiv  \int \frac{\mathrm{d}r}{f(r)}
\end{equation}
Finally, the effective potential barrier is given by
\begin{equation}
V_s(r) = f(r) \: \left(\frac{l (l+1)}{r^2}+\frac{f'(r)}{r} \right), \; \; \; \; l \geq 0
\end{equation}
where the prime denotes differentiation with respect to $r$, and $l$ is the angular degree.

\begin{figure*}[ht]
\centering
\includegraphics[width=0.32\textwidth]{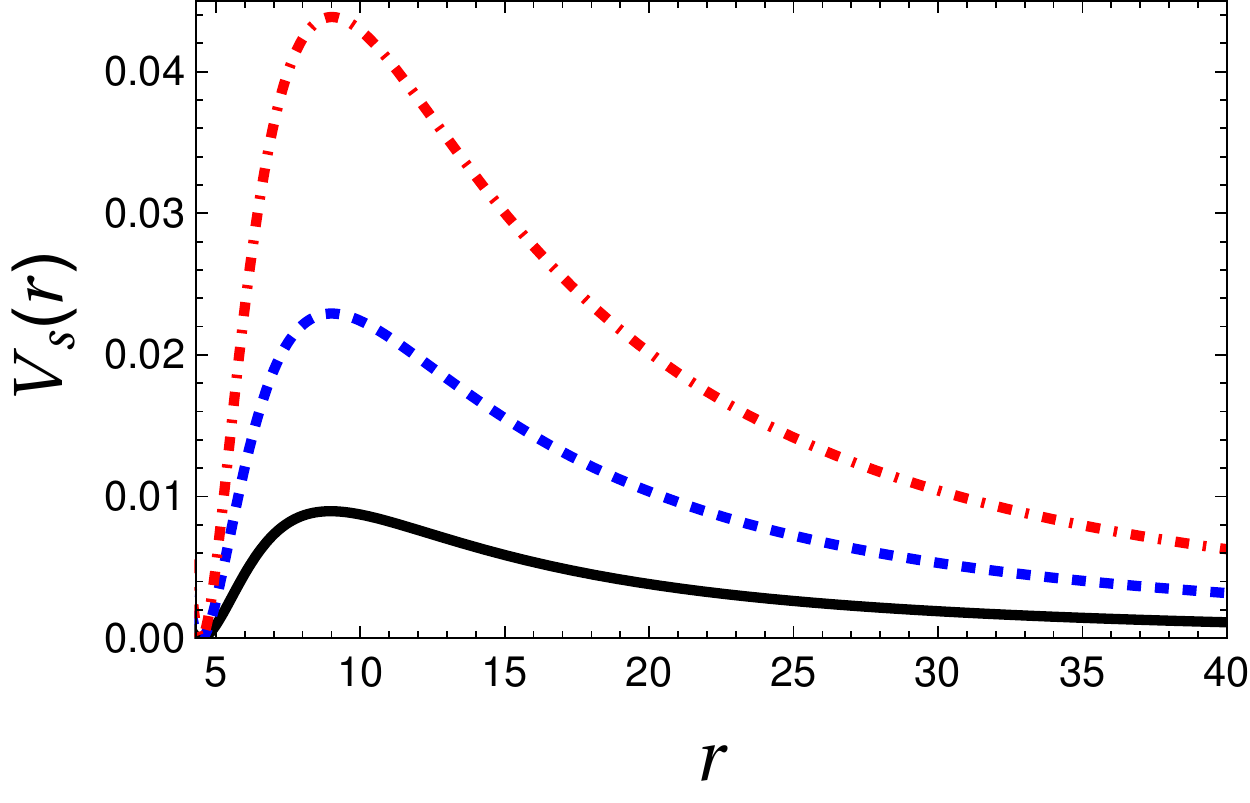}   \
\includegraphics[width=0.32\textwidth]{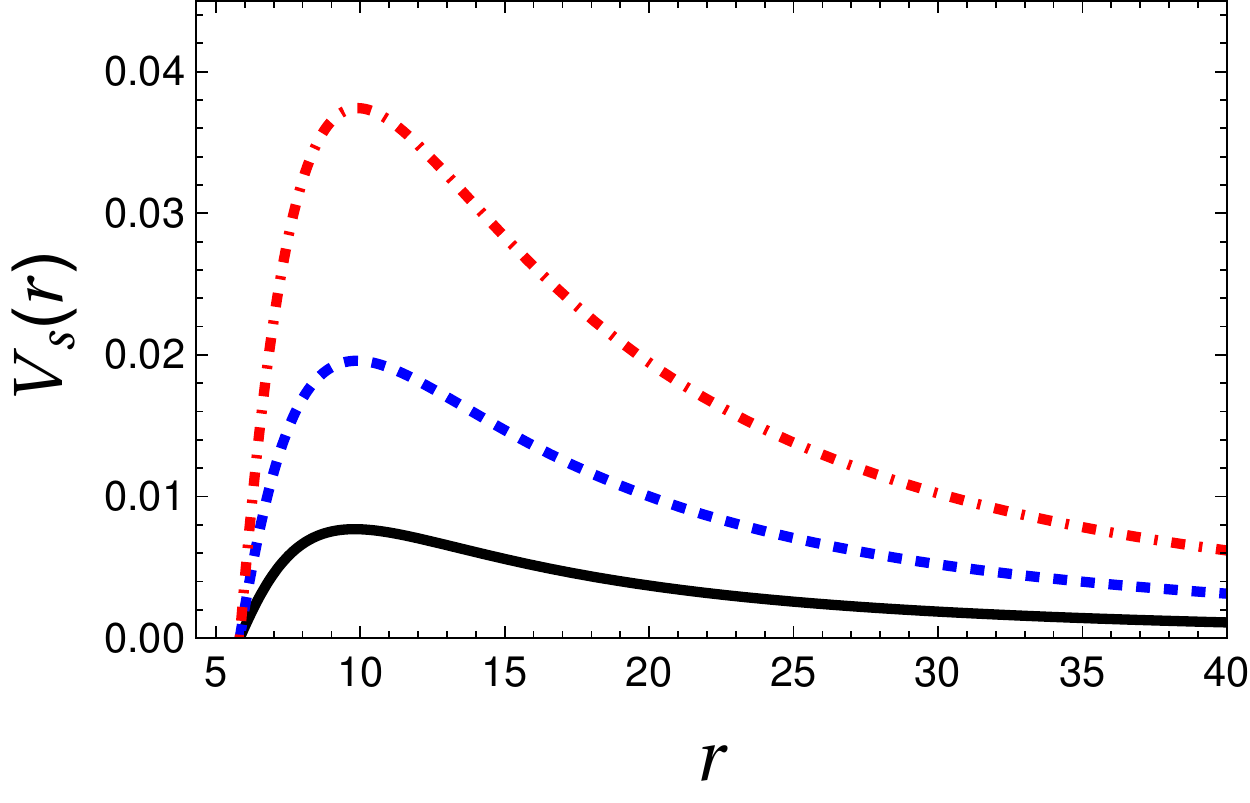}   \
\includegraphics[width=0.32\textwidth]{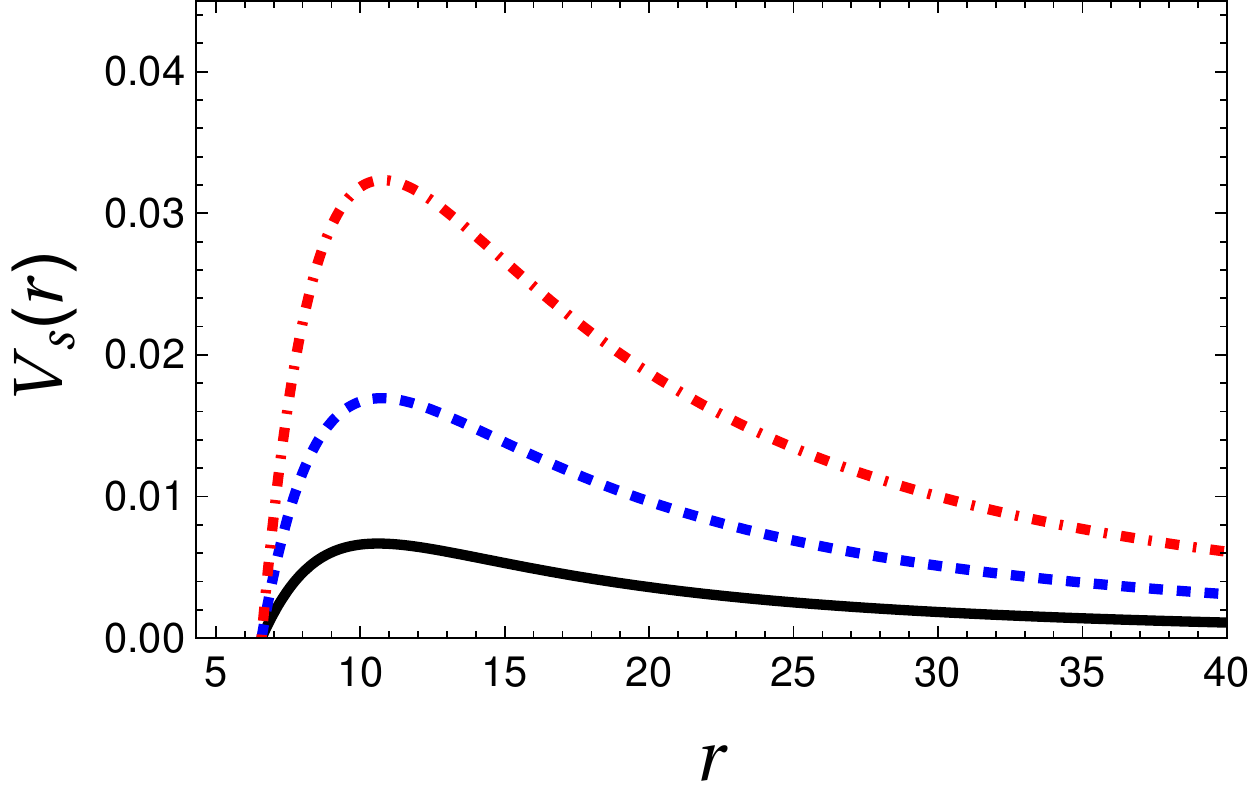}   \
\caption{
Effective potential barrier for scalar perturbations, $V_s(r)$, as a function of the radial coordinate $r$ assuming three different values of the parameter $M$. The panels in the first (left), second (center) and third (right) show $V_{s}(r)$ for: i) $M \approx 3.503$ (case 1), ii) $M=3.75$ (case 2) and iii) $M=4$ (case 3), respectively. In all three figures we have taken $\tilde{\omega} = 118/15\pi$ and $\gamma=9/2$. In each figure we show three different curves corresponding to: i) $l = 1$ (solid black line), ii) $l = 2$ (dashed blue line) and iii) $l = 3$ (dotted-dashed red line).
}
\label{fig:potential}
\end{figure*}

\begin{figure*}[ht]
\centering
\includegraphics[width=0.32\textwidth]{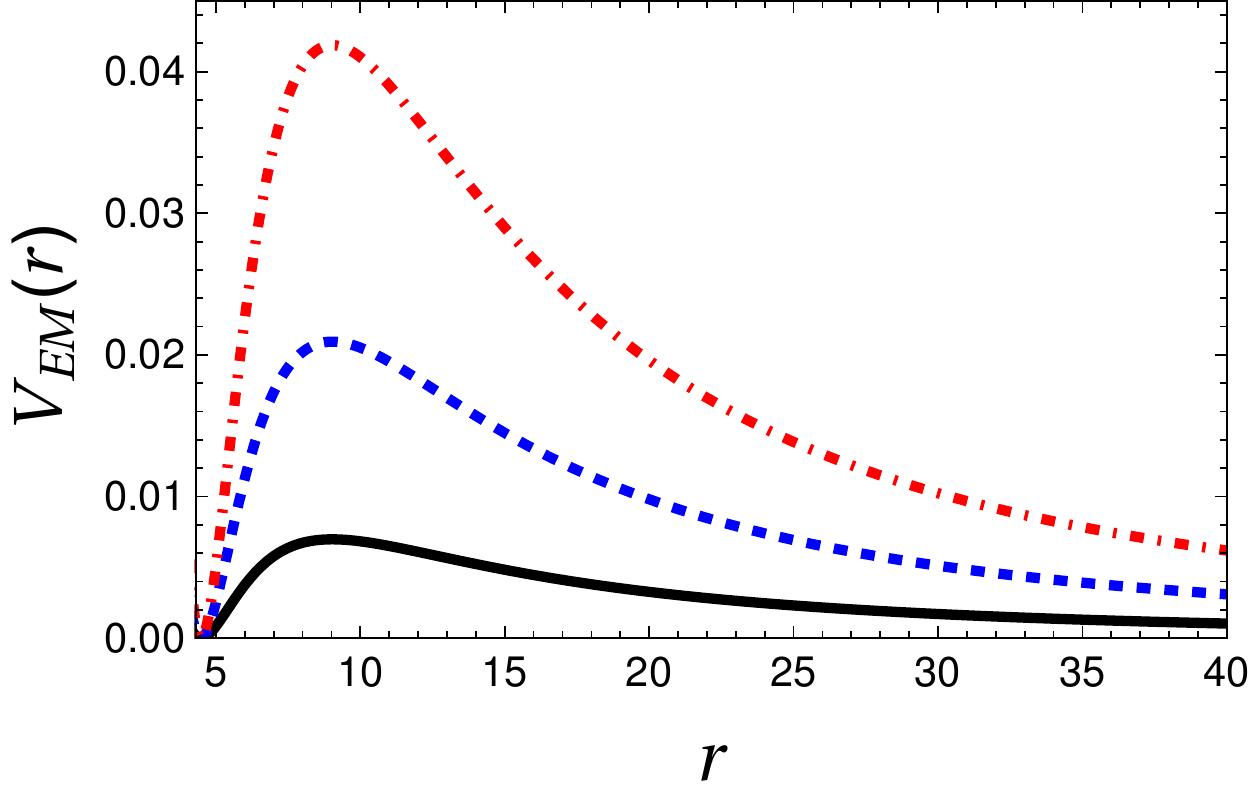}   \
\includegraphics[width=0.32\textwidth]{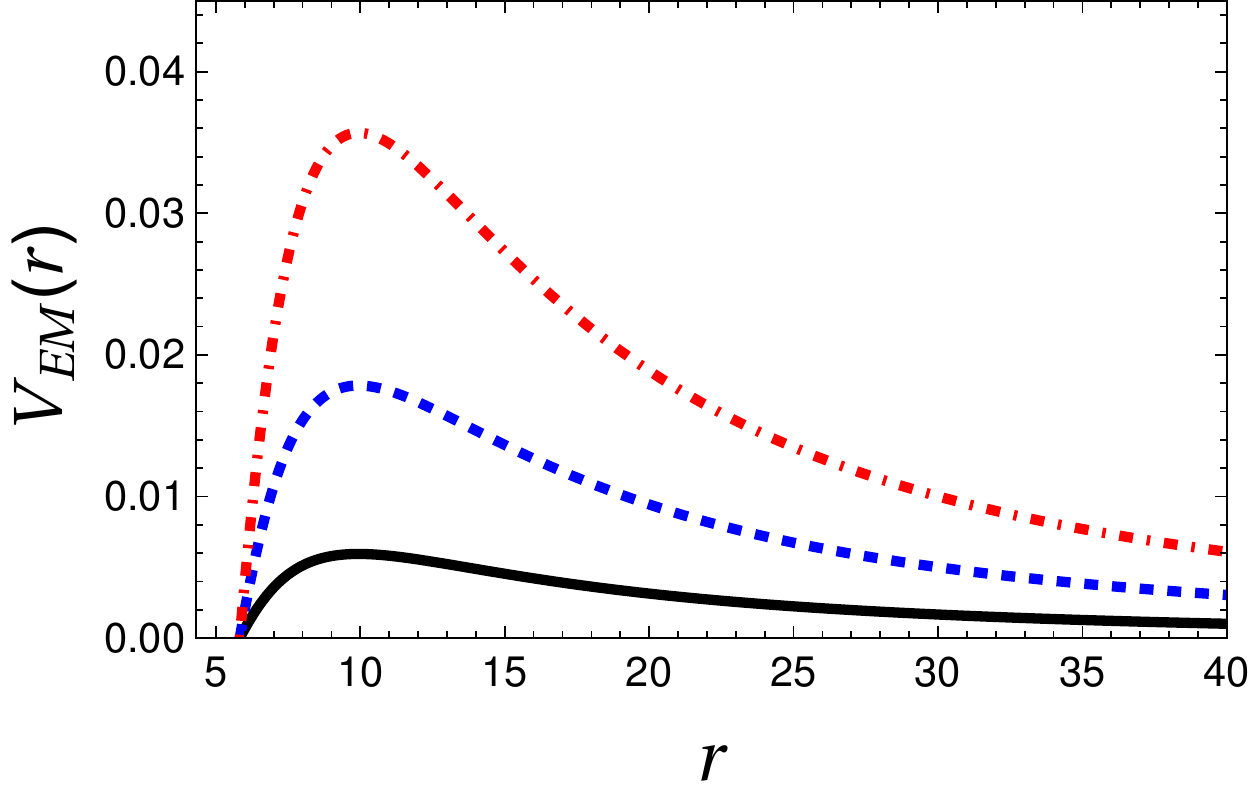}   \
\includegraphics[width=0.32\textwidth]{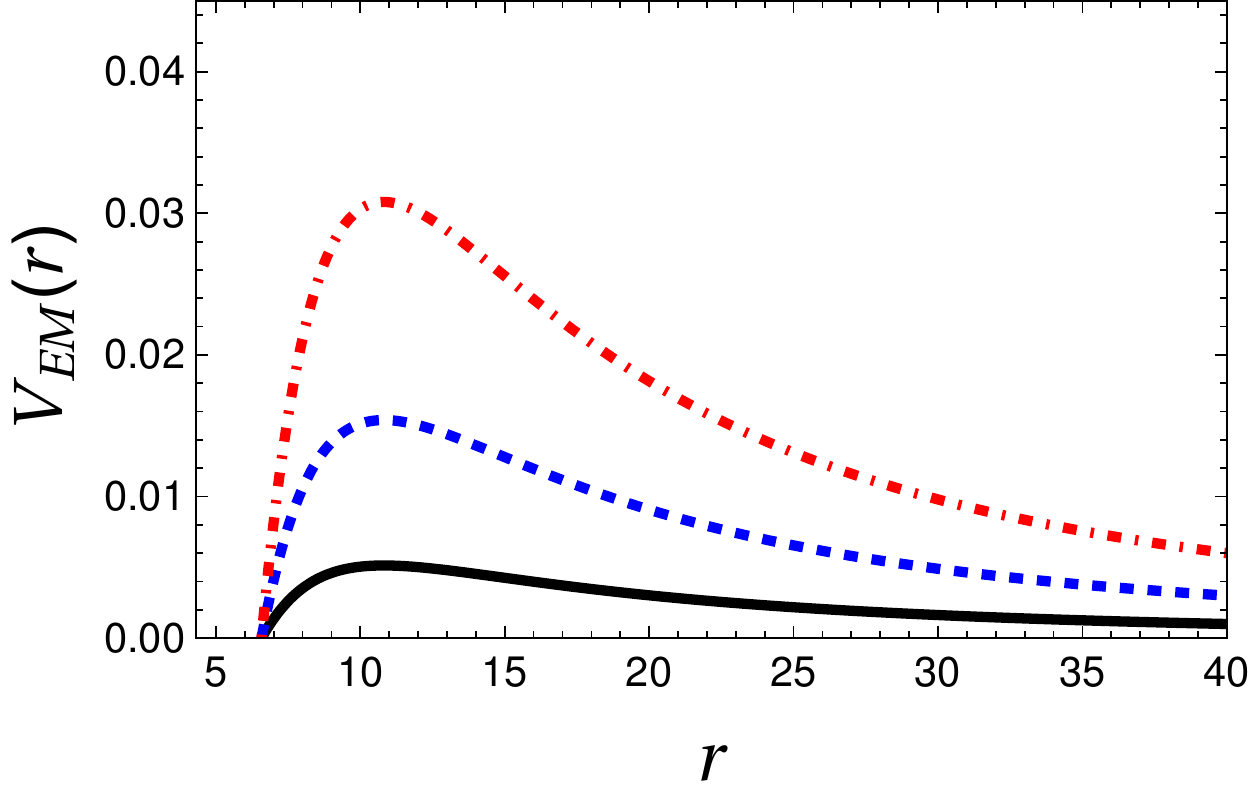}   \
\caption{
Same as in Fig.~\ref{fig:potential}, but for the effective potential barrier for electromagnetic perturbations, $V_{EM}(r)$. 
}
\label{fig:potential2}
\end{figure*}

Furthermore, electromagnetic perturbations are governed by Maxwell's equations \cite{Cardoso:2001bb}
\begin{equation}
F^{\mu \nu}_{;\nu} = 0, \; \; \; \; \; \; F_{\mu \nu} \equiv \partial_\mu A_\nu - \partial_\nu A_\mu
\end{equation}
where $A_\mu$ is the Maxwell potential, $F_{\mu \nu}$ is the electromagnetic field strength, and a semi-colon denotes covariant derivative. Following a similar separation of variables as before, one obtains the following effective potential barrier for electromagnetic perturbations \cite{lemos,Cardoso:2001bb}
\begin{equation}
V_{EM}(r) = f(r) \: \left( \frac{l (l+1)}{r^2} \right), \; \; \; \; l \geq 1
\end{equation}
Clearly, the effective potential vanishes at $r=r_H$ (when a horizon exists). What is more, for asymptotically flat space-times the effective potential barrier tends to zero when $r \rightarrow \infty$, while in between it reaches a maximum value $V_{\text{max}}$.
The effective potential barrier for scalar perturbations, $V_s(r)$, as a function of the radial coordinate $r$ is shown in the 3 panels of Fig.~\eqref{fig:potential}, while the effective potential barrier for electromagnetic perturbations, $V_{EM}(r)$, as a function of the radial coordinate $r$ is shown in the 3 panels of Fig.~\eqref{fig:potential2}. 
Each panel corresponds to a certain mass ($M \approx 3.503$ for left panel, $M=3.75$ for middle panel, and $M=4$ for right panel), while the 3 curves correspond to $l=1,2,3$.

Finally, the appropriate boundary conditions must be imposed. In the case of asymptotically flat space-times, the Schr{\"o}dinger-like equation is supplemented by the following boundary conditions \cite{valeria}
\begin{equation}
\psi(x) \rightarrow
\left\{
\begin{array}{lcl}
A e^{i \omega x} & \mbox{ if } & x \rightarrow - \infty \\
&
&
\\
 C_+ e^{i \omega x} + C_- e^{-i \omega x} & \mbox{ if } & x \rightarrow + \infty
\end{array}
\right.
\end{equation}
where the set of constants $\{A, C_{+}, C_{-}\}$ are arbitrary coefficients. 

To compute the QNMs, one imposes the quasinormal condition, i.e. $C_+ = 0$. 
That condition allows us to obtain an infinite set of discrete complex numbers, $\omega=\omega_R + i \omega_I $, called the quasinormal frequencies of the black hole. It is easy to verify that when $\omega_I > 0$ the perturbation grows exponentially (unstable mode), whereas when $\omega_I < 0$ the perturbation decays exponentially (stable mode). In the latter case, the real part determines the frequency of the oscillation, $f=\omega_R / (2 \pi)$, while the imaginary part determines the dumping time, $t_D = 1/|\omega_I|$.


\section{QNM of an improved Schwarzschild BH in the WKB approximation}

As usual in Physics, exact analytical calculations for QN spectra of black holes may be performed in very few cases, e.g. either when the effective potential barrier takes the form of the P{\"o}schl-Teller potential \cite{potential,ferrari,lemos,molina,panotop1}, or when the radial part of the wave function may be recast into the Gauss' hypergeometric function \cite{exact1,exact2,exact3,exact4,exact5,exact6}. More generically, over the years several different methods to compute the QNMs of black holes have been developed, such as the Frobenius method, generalization of the Frobenius series, fit and interpolation approach, method of continued fraction etc. For more details the interested reader may want to consult e.g. \cite{review3}. In particular, semi-analytical methods based on the Wentzel-Kramers-Brillouin (WKB) approximation (familiar from the standard non-relativistic quantum mechanics, see e.g. Gamow's work on the theory of $\alpha$-decay \cite{adecay} for a class of radioactive nuclei) \cite{wkb1,wkb2,wkb3} are among the most popular ones, and they have been extensively applied to several cases. For a partial list see for instance \cite{paper1,paper2,paper3,paper4,paper5,paper6}, and for more recent works \cite{paper7,paper8,paper9,paper10,Rincon:2018sgd}, and references therein.

Within the WKB approximation the QN frequencies are computed by
\begin{equation}
\omega_n^2 = V_0+(-2V_0'')^{1/2} \Lambda(n) - i \nu (-2V_0'')^{1/2} [1+\Omega(n)]
\end{equation}
where $n=0,1,2...$ is the overtone number, $\nu=n+1/2$, $V_0$ is the maximum of the effective potential, $V_0''$ is the second derivative of the effective potential evaluated at the maximum, while $\Lambda(n), \Omega(n)$ are complicated expressions of $\nu$ and higher derivatives of the potential evaluated at the maximum, and can be seen e.g. in \cite{paper2,paper7}. 
Here we have used the Wolfram Mathematica \cite{wolfram} code with WKB at any order from one to six presented in \cite{code}. For a given angular degree $l$ we have considered values $n \leq l$ only, since this is the case for which the best results are obtained, see e.g. Tables II, III, IV and V of \cite{Opala}.
For higher order WKB corrections, and recipes for simple, quick, efficient and accurate computations see \cite{Opala,Konoplya:2019hlu,RefExtra2}.


\begin{table*} 
\centering
\caption{QN frequencies (scalar perturbations) for $M=5, \gamma=9/2$, and $\tilde{\omega}=118/15\pi$. For comparison reasons, we also show two different cases: i) the standard one, i.e., $\tilde{\omega}=0$ (in parenthesis) and ii) the approximation computed in \cite{chinosII} with $\xi = 8.74185$ (in curly brackets).}
\resizebox{2\columnwidth}{!}{%
\begin{tabular}{cccccc}
\hline
$n$ &  $l=0$ & $l=1$ & $l=2$ & $l=3$ & $l=4$\\
\hline
     &   0.0241367\, -0.0181878 i   &   0.0604828\, -0.0184862 i   &   0.0997920\, -0.0182964 i   &   0.1393220\, -0.0182597 i   &   0.1789250\, -0.0182468 i  \\
0    &  (0.0220929\, -0.0201637 i)  &  (0.0585819\, -0.0195523 i)  &  (0.0967284\, -0.0193532 i)  &  (0.1350730\, -0.0193001 i)  &  (0.1734830\, -0.0192784 i) \\
     & \{0.0234430\, -0.0194976 i\} & \{0.0612570\, -0.0188642 i\} & \{0.1010160\, -0.0186862 i\} & \{0.1410110\, -0.0186411 i\} & \{0.1810820\, -0.0186233 i\}\\
\hline
     &                              &   0.0555291\, -0.0573753 i   &   0.0963757\, -0.0556588 i   &   0.1367950\, -0.0551772 i   &   0.1769310\, -0.0549829 i  \\
1    &                              &  (0.0528942\, -0.0613036 i)  &  (0.0927694\, -0.0591254 i)  &  (0.1321340\, -0.0584575 i)  &  (0.1711620\, -0.0581753 i) \\
     &                              & \{0.0565552\, -0.0586798 i\} & \{0.0977451\, -0.0569069 i\} & \{0.1385840\, -0.0563668 i\} & \{0.1791660\, -0.0561408 i\}\\
\hline
     &                              &                              &   0.0903912\, -0.0950511 i   &   0.1320580\, -0.0932217 i   &   0.1730940\, -0.0924225 i  \\
2    &                              &                              &  (0.0860772\, -0.1017400 i)  &  (0.1267180\, -0.0992021 i)  &  (0.1667370\, -0.0980645 i) \\ 
     &                              &                              & \{0.0921857\, -0.0973681 i\} & \{0.1341010\, -0.0953477 i\} & \{0.1755080\, -0.0944447 i\}\\
\hline
     &                              &                              &                              &   0.1256980\, -0.1329170 i    &   0.1677050\, -0.1309530 i  \\
3    &                              &                              &                              &  (0.1196860\, -0.1422770 i)   &  (0.1606370\, -0.1394960 i) \\
     &                              &                              &                              & \{0.1282510\, -0.1361580 i\}  & \{0.1704520\, -0.1339640 i\}\\
\hline
     &                              &                              &                              &                             &   0.1611590\, -0.1708380 i  \\
4    &                              &                              &                              &                             &  (0.1534430\, -0.1828460 i) \\
     &                              &                              &                              &                             & \{0.1644640\, -0.1749860 i\}\\
\hline     
\end{tabular}
\label{table:First_set}
}
\end{table*}



\begin{table*}
\centering
\caption{QN frequencies (scalar perturbations) for $M=8, \gamma=9/2$, and $\tilde{\omega}=118/15\pi$. For comparison reasons, we also show two different cases: i) the classical one, i.e., $\tilde{\omega}=0$ (in parenthesis) and ii) the approximation computed in \cite{chinosII} with $\xi = 8.33497$ (in curly brackets).}
\resizebox{2\columnwidth}{!}{%
\begin{tabular}{cccccc}
\hline
$n$ & $l=0$ & $l=1$ & $l=2$ & $l=3$ & $l=4$\\
\hline
     &  0.0141207\, -0.0125909 i    &   0.0370570\, -0.0120052 i   &   0.0611573\, -0.0118763 i   &   0.0853888\, -0.0118467 i   &   0.1096640\, -0.0118352 i   \\
0    & (0.0138080\, -0.0126024 i)   &  (0.0366137\, -0.0122202 i)  &  (0.0604552\, -0.0120958 i)  &  (0.0844208\, -0.0120626 i)  &  (0.1084270\, -0.0120490 i)  \\
     &\{0.0141135\, -0.0124528 i\}  & \{0.0371967\, -0.0120861 i\} & \{0.0613844\, -0.0119644 i\} & \{0.0857054\, -0.0119329 i\} & \{0.1100700\, -0.0119203 i\} \\
\hline
     &                              &   0.0337423\, -0.0375315 i   &   0.0588411\, -0.0362336 i   &   0.0836685\, -0.0358552 i   &   0.1083050\, -0.0356977 i   \\
1    &                              &  (0.0330589\, -0.0383148 i)  &  (0.0579808\, -0.0369534 i)  &  (0.0825839\, -0.0365360 i)  &  (0.1069760\, -0.0363596 i)  \\
     &                              & \{0.0338768\, -0.0377944 i\} & \{0.0590699\, -0.0365137 i\} & \{0.0839865\, -0.0361232 i\} & \{0.1087120\, -0.0359589 i\} \\
\hline
     &                              &                              &  0.0549027\, -0.0621947 i    &   0.0804891\, -0.0607583 i   &   0.1057100\, -0.0601196 i   \\
2    &                              &                              & (0.0537982\, -0.0635875 i)   &  (0.0791989\, -0.0620013 i)  &  (0.1042110\, -0.0612903 i)  \\ 
     &                              &                              &\{0.0551577\, -0.0627104 i\}  & \{0.0808193\, -0.0612348 i\} & \{0.1061240\, -0.060574 i\}  \\
\hline
     &                              &                              &                              &   0.0763352\, -0.0869715 i   &   0.1021210\, -0.0854093 i   \\
3    &                              &                              &                              &  (0.0748036\, -0.0889231 i)  &  (0.1003980\, -0.0871852 i)  \\
     &                              &                              &                              & \{0.0767066\, -0.0876947 i\} & \{0.1025560\, -0.0860834 i\} \\
\hline 
     &                              &                              &                              &                              &   0.0978641\, -0.1117760 i   \\
4    &                              &                              &                              &                              &  (0.0959021\, -0.1142780 i)  \\
     &                              &                              &                              &                              & \{0.0983489\, -0.1127010 i\} \\
\hline 
\end{tabular}
\label{table:Second_set}
}
\end{table*}



\begin{table*}
\centering
\caption{QN frequencies (electromagnetic perturbations) for $M=5, \gamma=9/2$, and $\tilde{\omega}=118/15\pi$. For comparison reasons, we also show two different cases: i) the classical one, i.e., $\tilde{\omega}=0$ (in parenthesis) and ii) the approximation computed in \cite{Li:2013kkb} with $\xi = 8.74185$ (in curly brackets).}
\begin{tabular}{ccccc}
\hline
$n$ &  $l=1$ & $l=2$ & $l=3$ & $l=4$ \\
\hline
     &   0.0518962\, -0.0176598 i    &   0.0947619\, -0.0179553 i   &   0.1357420\, -0.0180840 i    &   0.1761450\, -0.0181403 i  \\
0    &  (0.0496383\, -0.0185274 i)   &  (0.0915187\, -0.0190022 i)  &  (0.1313800\, -0.0191234 i)   &  (0.1706190\, -0.0191720 i) \\
     & \{0.0525392\, -0.0180371 i\}  & \{0.0959220\, -0.0183708 i\} & \{0.1373920\, -0.0184799 i\}  & \{0.1782740\, -0.0185257 i\} \\
\hline
     &   0.0464157\, -0.0555774 i    &   0.0912135\, -0.0546942 i   &   0.1331670\, -0.0546641 i    &   0.1741290\, -0.0546682 i  \\
1    &  (0.0428590\, -0.0588237 i)   &  (0.0873068\, -0.0581456 i)  &  (0.1283470\, -0.0579461 i)   &  (0.1682530\, -0.0578630  i) \\
     & \{0.0470723\, -0.0567635 i\}  & \{0.0924695\, -0.0560231 i\} & \{0.1348990\, -0.0558988 i\}  & \{0.1763260\, -0.0558538 i\} \\
\hline
     &                               &   0.0850381\, -0.0936163 i   &   0.1283470\, -0.0924114 i    &   0.1702500\, -0.0919144 i  \\
2    &                               &  (0.0801812\, -0.1003460 i)  &  (0.1227580\, -0.0984128 i)   &  (0.1637440\, -0.0975669 i) \\ 
     &                               & \{0.0866187\, -0.0960839 i\} & \{0.1302960\, -0.0946181 i\}  & \{0.1726090\, -0.0939850 i\} \\
\hline
     &                               &                              &   0.1218910\, -0.1318680 i    &   0.1648060\, -0.1302740 i  \\
3    &                               &                              &  (0.1154960\, -0.1413000 i)   &  (0.1575260\, -0.1388480 i) \\
     &                               &                              & \{0.1242960\, -0.1352350 i\}  & \{0.1674740\, -0.1333580 i\} \\
\hline
     &                               &                              &                              &   0.1582030\, -0.1700180 i  \\
4    &                               &                              &                              &  (0.1501910\, -0.1820910 i) \\
     &                               &                              &                              & \{0.1613950\, -0.1742670 i\} \\
\hline
\end{tabular}
\label{table:Third_set}
\end{table*}



\begin{table*}
\centering
\caption{QN frequencies (electromagnetic perturbations) for $M=8, \gamma=9/2$, and $\tilde{\omega}=118/15\pi$. For comparison reasons, we also show two different cases: i) the classical one, i.e., $\tilde{\omega}=0$ (in parenthesis) and ii) the approximation computed in \cite{Li:2013kkb} with $\xi = 8.33457$ (in curly brackets).}
\begin{tabular}{ccccc}
\hline
$n$ &  $l=1$ & $l=2$ & $l=3$ & $l=4$\\
\hline
     &   0.0315513\, -0.0114118 i   &   0.0579389\, -0.0116614 i   &   0.0831040\, -0.0117377 i    &   0.1078920\, -0.0117693 i  \\
0    &  (0.0310239\, -0.0115796 i)  &  (0.0571992\, -0.0118764 i)  &  (0.0821123\, -0.0119521 i)   &  (0.1066370\, -0.0119825 i) \\
     & \{0.0316524\, -0.0114931 i\} & \{0.0581510\, -0.0117535 i\} & \{0.0834114\, -0.0118262 i\}  & \{0.1082910\, -0.0118559 i\} \\
\hline
     &   0.0276887\, -0.0361299 i   &   0.0554914\, -0.0356318 i   &   0.0813353\, -0.0355388 i    &   0.1065100\, -0.0355041 i  \\
1    &  (0.0267869\, -0.0367648 i)  &  (0.0545667\, -0.0363410 i)  &  (0.0802171\, -0.0362163 i)   &  (0.1051580\, -0.0361644 i) \\
     & \{0.0277253\, -0.0363836 i\} & \{0.0556936\, -0.0359246 i\} & \{0.0816398\, -0.0358140 i\}  & \{0.1069080\, -0.0357698 i\} \\
\hline
     &                              &   0.0513362\, -0.0613273 i   &   0.0780673\, -0.0602670 i    &   0.1038710\, -0.0598101 i  \\
2    &                              &  (0.0501132\, -0.0627160 i)  &  (0.0767235\, -0.0615080 i)   &  (0.1023400\, -0.0609793 i) \\
     &                              & \{0.0515419\, -0.0618669 i\} & \{0.0783756\, -0.0607563 i\}  & \{0.1042720\, -0.0602723 i\} \\ 
\hline
     &                              &                              &   0.0738002\, -0.0863555 i    &   0.1002220\, -0.0850032 i  \\
3    &                              &                              &  (0.0721851\, -0.0883122 i)   &  (0.0984539\, -0.0867798 i) \\
     &                              &                              & \{0.0741372\, -0.0870981 i\}  & \{0.1006390\, -0.0856887 i\} \\
\hline
     &                              &                              &                               &   0.0958965\, -0.1112980 i  \\
4    &                              &                              &                               &  (0.0938697\, -0.1138070 i) \\
     &                              &                              &                               & \{0.0963548\, -0.1122390 i\} \\
\hline
\end{tabular}
\label{table:Fourth_set}
\end{table*}

\begin{figure*}[ht]
\centering
\includegraphics[width=0.48\textwidth]{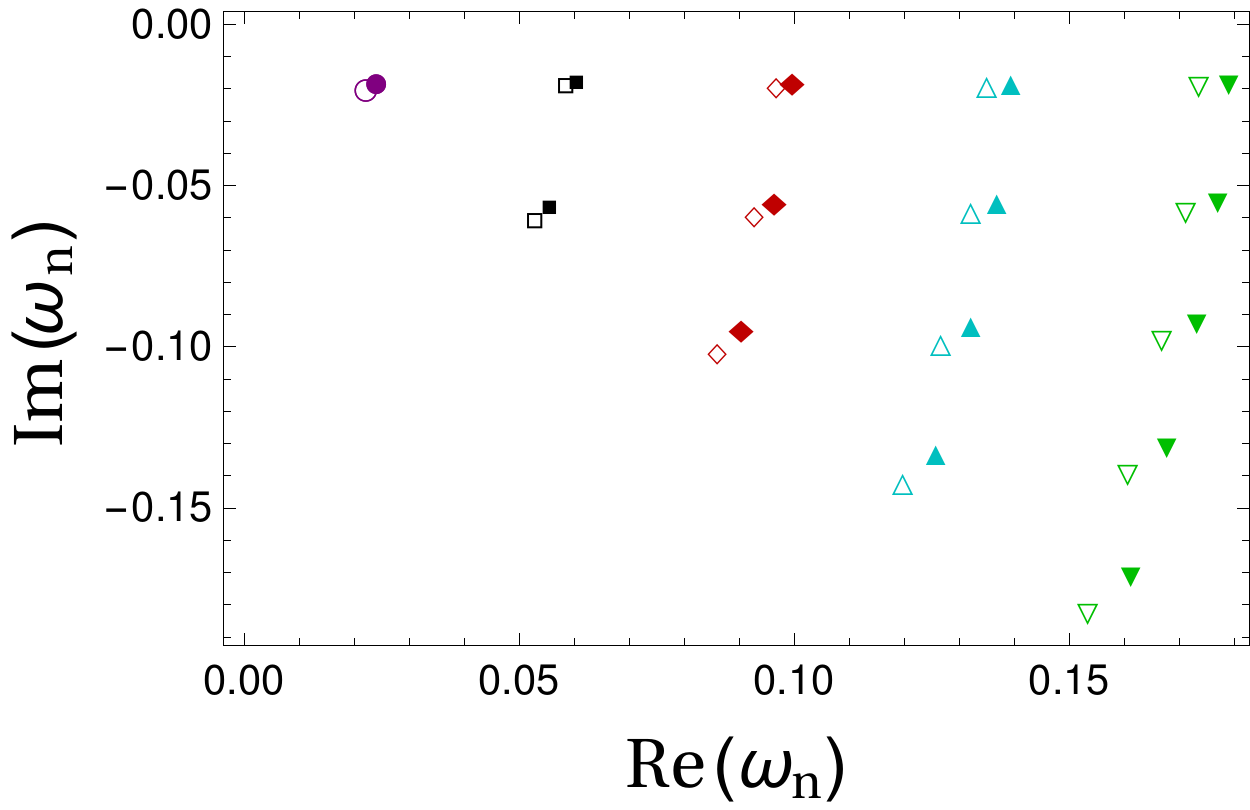}   
\ \ \
\includegraphics[width=0.48\textwidth]{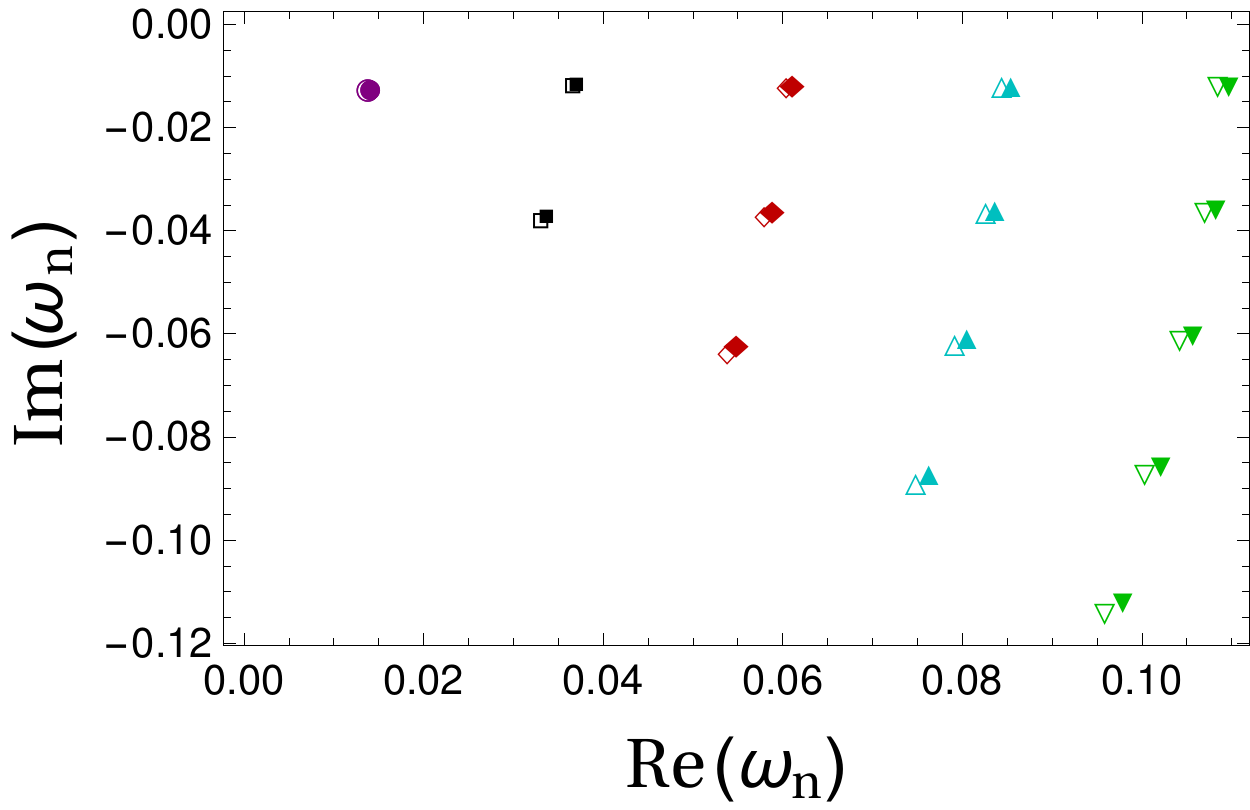}   

\caption{
Quasinormal modes (scalar perturbations) $\omega_n \equiv \omega_R + i \omega_I$. 
{\bf{LEFT:}} $\text{Im}(\omega_n)$ versus $\text{Re}(\omega_n)$ for $M=5$ in five different cases: 
i)   $l=0$ (purple point), 
ii)  $l=1$ (black square), 
iii) $l=2$ (red rhombus), 
iv)  $l=3$ (cyan triangle), and
 v)  $l=4$ (green inverted triangle)
{\bf{RIGHT:}} Same as in the left panel, but for $M=8$. Empty point, squares, rhombuses and triangles correspond to the modes of the classical geometry.
}
\label{fig:frequencies_scalar}
\end{figure*}
\begin{figure*}[ht]
\centering
\includegraphics[width=0.48\textwidth]{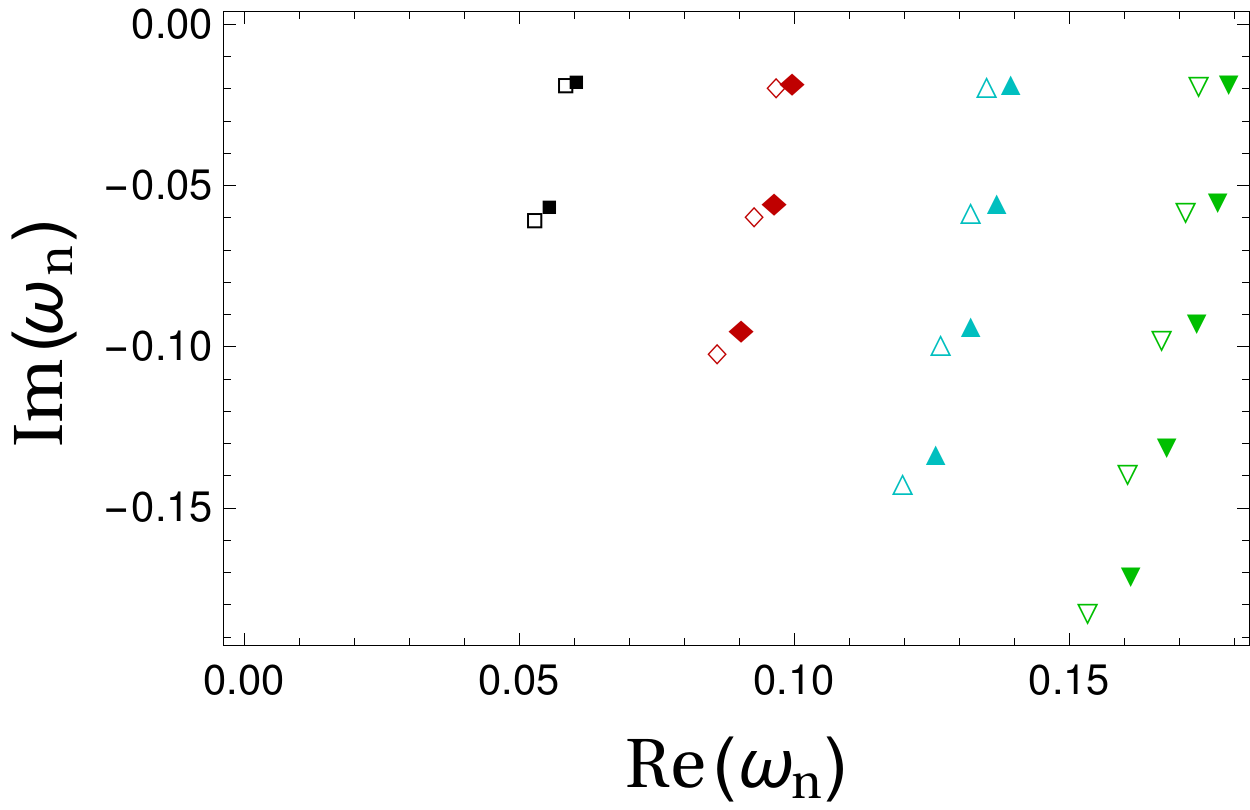}   
\ \ \
\includegraphics[width=0.48\textwidth]{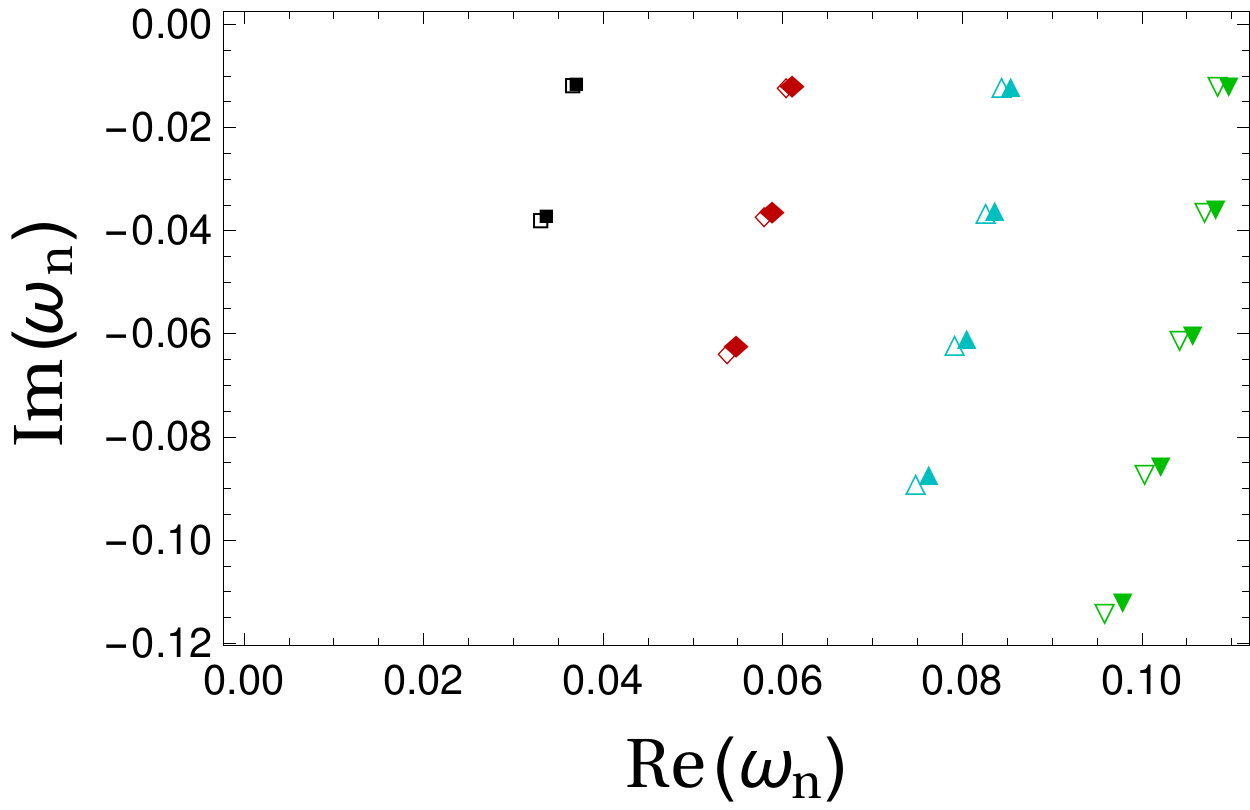}   

\caption{
Quasinormal modes (electromagnetic perturbations) $\omega_n \equiv \omega_R + i \omega_I$. 
{\bf{LEFT:}} $\text{Im}(\omega_n)$ versus $\text{Re}(\omega_n)$ for $M=5$ in four different cases: 
i)   $l=1$ (purple point), 
ii)  $l=2$ (black triangle), 
iii) $l=3$ (red rhombus) and 
iv)  $l=4$ (cyan square).
{\bf{RIGHT:}} Same as in left panel, but for $M=8$. Empty point, squares, rhombuses and triangles correspond to the the modes of the classical geometry. 
}
\label{fig:frequencies_EM}
\end{figure*}

The obtained QNMs are summarized in Tables \ref{table:First_set}, \ref{table:Second_set}, \ref{table:Third_set} and \ref{table:Fourth_set}, and they are pictorially shown in the Figures~ \ref{fig:frequencies_scalar} and \ref{fig:frequencies_EM}. We recall that the information on the stability of the modes is encoded into the sign of the imaginary part of the frequencies. see the end of last section. In particular, for stable modes $\omega_I < 0$, while for unstable modes $\omega_I > 0$, see the end of last section. Therefore all modes are found to be stable. Our results show that i) when the mass of the black hole increases both the real part and the absolute value of the imaginary part are reduced, ii) for a given mass $M$ and overtone number $n$, the real part of the modes increases with the angular degree, while the absolute value of the imaginary part decreases with $l$ in the case of scalar perturbations and increases with $l$ in the case of electromagnetic perturbations, and iii) for a given black hole mass and angular degree, the real part of the frequencies decreases with $n$, while the absolute value of the imaginary part of the modes increases with the overtone number.

\smallskip

For comparison reasons, the QN frequencies of the classical Schwarzschild space time as well as the QNMs of an improved black hole studied in \cite{chinosII} using a different approach are also shown in the tables. This demonstrates that a different identification for the cut-off does not influence the numerical values of the QN frequencies significantly. Regarding the real part, it is found to be higher than the one of the classical black hole, but lower than the real part obtained in \cite{chinosII} using a different approach. Regarding the imaginary part, its absolute value is found to be lower than the other two. 

\smallskip

The difference between the results for the classical BH and the improved BH considered here is more notorious in the higher excited modes and for $M=5$. To be more precise, for the fundamental mode of $l=2,3,4$ we find a difference of $(3-4) \%$. Although the lapse functions and Newton's constants are very different at short scales, the effective potentials exhibit very similar shapes (see Fig.~ \ref{fig:report}), which is why the numerical values of the frequencies are not dramatically modified. The reader should keep in mind, however, that this difference of a few percentage is observed only because the objects considered in the present work are hypothetical BHs with masses of the order of the Planck mass (specifically 5 and 8 Planck masses), rather than astrophysical BHs. It should be emphasized at this point that a very similar computation for more realistic, astrophysical BHs with a mass $M=10~M_{\odot}$, with $M_{\odot} =1.988 \times 10^{30}~$ kgr being the solar mass, shows that there is absolutely no difference between the frequencies of the classical BH and the frequencies of the improved BH studied here. Therefore, although the detection of $\sim 50$ gravitational wave events by the aLIGO/Virgo observatories is expected to produce accurate enough frequency results during the waveform ring-down phase \cite{Manfredi:2017xcv}, those results will not have the potential neither to distinguish between the two kinds of black holes nor to detect quantum effects.

\smallskip

Regarding future work, it would be both challenging and interesting to compute the QNMs for gravitational perturbations, since they may serve as a distinguisher between GR and alternative theories of gravity, see e.g. \cite{Bhattacharyya:2017tyc}. There it was shown that contrary to what happens in GR, where axial and polar modes share equal amounts of radiation, this does not hold in $f(R)$ theories of gravity. We hope to be able to address that issue in a forthcoming article.


\section{Conclusions}

Summarizing our work, in the present article we have studied the quasinormal spectrum for scalar as well as for electromagnetic perturbations of an improved Schwarzschild black hole. After discussing the main properties of the space-time, we perturbed the black hole with a test massless scalar/vector field, and we investigated its propagation into a fixed gravitational background. The wave equation with the corresponding effective potential barrier was obtained. Imposing the appropriate boundary conditions, the QN frequencies were numerically computed employing the WKB method of 6th order. For better visualization, we showed on the ($\omega_R-\omega_I$) plane the impact on the spectrum of the overtone number, the angular degree and the mass of the black hole. All modes were found to be stable. 
For comparison reasons we have shown in the same table the frequencies corresponding to i) the classical Schwarzschild solution, and ii) to the modes obtained in a previous work using a different approach. Our findings show that i) when a different cut-off identification is made the results for the spectra do change, as expected, although not significantly, and ii) for hypothetical objects with masses comparable to the Planck mass, the difference in the numerical values between the modes of the classical solution and the modes of the improved solution studied here is of the order of a few per cent, whereas for realistic, astrophysical BHs no difference in the frequencies was observed.



\section*{Acknowlegements}

We are grateful to the reviewer for a careful reading of the manuscript, 
for a constructive criticism as well as for useful comments and suggestions 
that helped us improve the presentation and quality of our work significantly.  
The author \'A.~R. acknowledges DI-VRIEA for financial support through Proyecto 
Postdoctorado 2019 VRIEA-PUCV. The author G.~P. thanks the Fun\-da\c c\~ao para 
a Ci\^encia e Tecnologia (FCT), Portugal, for the financial support to the Center 
for Astrophysics and Gravitation-CENTRA, Instituto Superior T\'ecnico, Universidade 
de Lisboa, through the Project No.~UIDB/00099/2020.


\bibliographystyle{unsrt} 
\bibliography{Bibliography}  

\end{document}